\documentclass[letterpaper,10pt,twocolumn,final,journal,oneside]{IEEEtran}

\usepackage{textcomp}
\makeatletter
\def\endtable{\end@float}
\def\endfigure{\end@float}
\makeatother
\usepackage{tabularx}
\usepackage{cite}
\usepackage{array}
\usepackage{graphicx}
\usepackage[cmex10]{amsmath}
\usepackage{mathtools}
\usepackage{multirow}
\usepackage{gensymb}
\usepackage{nomencl}
\usepackage{color}
\usepackage{float}
\usepackage{balance}
\usepackage{amssymb}
\usepackage{hyperref}
\usepackage{colortbl}
\usepackage{tabularray}
\usepackage{ctable}
\usepackage{notoccite}
\usepackage{acronym}
\usepackage{enumitem}
\usepackage{comment}

\usepackage{flushend}
\usepackage{algorithm}
\usepackage[noend]{algpseudocode}

\newcommand{\mr}{\mathrm}

\usepackage{dcolumn}
\newcolumntype{d}[1]{D{.}{.}{#1}}

\newcolumntype{L}[1]{>{\raggedright\let\newline\\\arraybackslash\hspace{0pt}}m{#1}}
\newcolumntype{C}[1]{>{\centering\let\newline\\\arraybackslash\hspace{0pt}}m{#1}}
\newcolumntype{R}[1]{>{\raggedleft\let\newline\\\arraybackslash\hspace{0pt}}m{#1}}

\acrodef{der}[DER]{distributed energy resource}
\acrodef{res}[RES]{renewable energy source}
\acrodef{sg}[SG]{synchronous generator}
\acrodef{ibr}[IBR]{inverter-based resource}
\acrodef{pll}[PLL]{phase-locked loop}
\acrodef{avr}[AVR]{automatic voltage regulator}
\acrodef{pcc}[PCC]{point of common coupling}
\acrodef{svm}[SVM]{support vector machine}
\acrodef{pi}[PI]{proportional-integral}
\acrodef{asm}[ASM]{adaptive sampling method}
\acrodef{scr}[SCR]{short-circuit ratio}
\acrodef{rpi}[RPI]{region of practical interest}
\acrodef{tso}[TSO]{Transmission System Operator}
\graphicspath{{Figs/}}

\begin{document}

\markboth{A Manuscript Submitted to the IEEE Transactions on Power Systems}{}
\title{Small-Signal Stability Manifolds in Converter-Dominated Power Systems}

\author{%
        Francesco Conte,~\IEEEmembership{Senior Member,~IEEE,} 
        Fernando Mancilla-David,~\IEEEmembership{Member,~IEEE,}\\ Federico Silvestro,~\IEEEmembership{Senior Member,~IEEE,} Samuele~Grillo,~\IEEEmembership{Senior Member,~IEEE}%
        \thanks{This work has been funded by the EU fund Next Generation EU, Missione 4, Componente 1, CUP D53D23001650006, MUR PRIN project 2022ZJPPSN SCooPS.}%
        \thanks{F. Conte is with FDI, Università Campus Bio-Medico di Roma, Rome, Italy. (email: f.conte@unicampus.it).}%
        \thanks{F. Mancilla-David is with the EE Department, University of Colorado Denver, Denver, Colorado, USA. (email: fernando.mancilla-david@ucdenver.edu).}%
        \thanks{F. Silvestro is with DITEN, Università degli Studi di Genova, Genoa, Italy. (email: federico.silvestro@unige.it).}%
        \thanks{S. Grillo is with DEIB, %
        Politecnico di Milano, Milano, Italy %
        (e-mail: samuele.grillo@polimi.it).}%
}

\maketitle

\begin{abstract}
This paper proposes a systematic framework to assess the small-signal stability of power systems with high shares of grid-following \acp{ibr} under varying controller parameters and operating conditions. Stability manifolds are introduced to identify controller-parameter regions that ensure stability across multiple scenarios. Full-network linearization and eigenvalue analysis are combined with adaptive sampling based on probabilistic support vector machine classification to approximate stability boundaries efficiently, while surrogate optimization identifies feasible initial controller settings meeting bandwidth and phase-margin constraints. The approach is validated on a modified Cigré European HV network benchmark with 50 operating scenarios and increasing inverter penetration. Results show that stability sensitivity grows with inverter share, interactions among \acp{ibr} reshape admissible parameter regions, and simplified equivalent-network models may overlook critical system-level limitations. The framework supports stability-oriented controller design and interconnection studies in converter-dominated systems.
\end{abstract}
\acresetall


\IEEEpeerreviewmaketitle

\section{Introduction}
\label{sec:intro}
Mainly due to the rapid expansion of renewables, the power system is transitioning from being predominantly based on \acp{sg} to one with high penetration of \acp{ibr}. In this new context, ensuring and analyzing stability requires assumptions and methodologies that differ from those in classical approaches, as they involve dynamic phenomena previously negligible or absent \cite{Hatziargyriou2021}. These include, for instance, the interactions between the control schemes of \acp{ibr} and other power system components.

\subsection{Literature review}

Most of the literature analyzes the stability of \acp{ibr} using a Thévenin-equivalent representation of the grid. This approach is widely employed to study the small-signal stability of grid-following \acp{ibr} and to assess the impact of control parameters under varying grid-strength conditions. For example, \cite{Rodriguez2019,Collados2020} examine the effects of control tuning, dead-time, and delays on converters connected to a Thévenin-equivalent grid. Other studies focus on the role of \ac{pll} dynamics in weak grids, typically modeled with high equivalent impedance \cite{Huang2022,Zhang2024}. Further small-signal analyses based on this modeling approach are reported in \cite{Saleem2023,Saleem2024,Ma2024}, and similar assumptions are also adopted for grid-forming \acp{ibr} \cite{Mohammed2024}. The widespread use of Thévenin-equivalent models has clarified the relationship between inverter stability and control parameters as a function of grid impedance. However, this approach is inherently limited to single-point interconnection studies and cannot capture interactions among multiple \acp{ibr} connected to the same network or between \acp{ibr} and \acp{sg}. Consequently, system-level effects related to network topology, generation mix, and multi-device interactions remain largely unexplored.

To address these limitations, recent studies have adopted more detailed network models for small-signal stability analysis. In \cite{Collados2022} and \cite{Ding2025}, full-network representations are used to analyze systems with multiple \acp{ibr}, capturing interactions beyond simplified equivalents. However, these methods entail high computational complexity, requiring repeated linearization of large-scale systems and eigenvalue analysis across multiple operating points and parameter settings, which limits their suitability for extensive parametric studies and controller tuning. In parallel, several studies analyze the impact of \ac{ibr} control strategies on power system stability through modal analysis, typically focusing on single operating points or simplified network equivalents \cite{Mourouvin:2021,Gu:2023}. Although they offer useful insights into the sensitivity of dominant modes to control parameters, they generally lack a systematic assessment of stability robustness across diverse operating conditions and network configurations. A growing body of literature has also questioned the validity of conventional reduced-order small-signal models in converter-dominated systems. Recent contributions have shown that neglecting fast electromagnetic dynamics of the network and \ac{sg} stator flux may lead to inaccurate stability assessments. In particular, \cite{Kelada:2025} proposes a detailed small-signal modeling framework that includes network dynamics and \ac{sg} stator flux dynamics, demonstrating that fast inner current control loops of grid-forming converters can interact with electromagnetic modes of machines and transmission lines. 

While the above studies provide valuable insights into modeling accuracy, control interactions, and fast dynamic phenomena, most existing works analyze stability at specific operating points or along limited parameter trajectories. As a result, a systematic characterization of stability robustness with respect to variations in \ac{ibr} control parameters across multiple operating conditions and network scenarios is still lacking.

\subsection{Paper contributions}

This paper \emph{contributes to understanding small-signal stability of power systems with increasing penetration of grid-following \acp{ibr}}. Stability is characterized through \emph{stability manifolds}, defined as regions of the controller-parameter space that preserve stability under multiple operating conditions. A comprehensive \emph{small-signal modeling and linearization tool} integrates \acp{sg}, \acp{ibr}, grid elements, and controls into a unified state-space model, with stability assessed via eigenvalue analysis across multiple operating conditions. To reduce the computational burden of exhaustive scans, an \emph{adaptive sampling method based on probabilistic \ac{svm} classification} identifies stability boundaries and evaluates sensitivity to key control parameters, including \ac{pll}, current, and dc-link voltage control gains. A \emph{surrogate-optimization-based} ~\cite{Jones:1998,matlab_surrogateopt} \emph{tuning procedure} is also proposed, combining classical design criteria with eigenvalue-based assessments. The framework is validated on a modified Cigré European HV benchmark~\cite{cigre} with different operating conditions and inverter penetration levels. Results show increasing sensitivity to controller tuning and significant interactions among inverter control loops. The methodology provides both analytical insight and practical support for \acp{tso} and \acp{ibr} owner for controller design, interconnection studies, and stability-oriented grid-code development in converter-dominated systems. Preliminary results obtained with the same appraoch have been presented in \cite{Conte:2025}.

The remainder of the paper is organized as follows. Section~\ref{sec:modeling_considerations} introduces the power system modeling assumptions; Section~\ref{sec:stability_analysis_tool} describes the small-signal stability analysis tool; Section~\ref{sec:stability_manifolds} introduces the method to identify the parameters stability manifolds; Section~\ref{sec:optimal_gains} presents the parameters optimal tuning method; Section~\ref{sec:case_study} introduces the modified Cigré European HV case study and present the results obtained through the proposed method; Section~\ref{sec:Conclusions} reports the concluding remarks.

\section{Modeling considerations}
\label{sec:modeling_considerations}
Devices that play a role in the small-signal stability of contemporary power systems include traditional equipment, such as \acp{sg} and their associated control systems, transmission lines, power transformers, reactive-element banks, and various types of load, as well as modern \ac{ibr} power generating units. In addition, the power system operating point, which is of central importance in small-signal stability studies, is strongly influenced by load variations and generation dispatch decisions. This section discusses the modeling assumptions adopted for both conventional and inverter-based equipment, as well as the synthesis of steady-state operating points used for linearization and stability analysis.

\subsection{Traditional equipment modeling and control}

\subsubsection{Synchronous generators}
The \acp{sg} are modeled using a classical sixth-order synchronous machine model \cite{Kundur1994,AndersonFouad2003}  formulated in the rotating $dq$ reference frame. This model captures the electromechanical rotor dynamics together with the transient and subtransient electromagnetic effects associated with the field and damper windings on both the $d$ and $q$ axes. 
The mechanical input power of each generator is regulated through a steam turbine and governor model operating with a $5\%$ steady-state frequency droop characteristic \cite{Kundur1994}. Voltage regulation is achieved by means of a standard IEEE Type~1 automatic voltage regulator (AVR) \cite{avrIEEE}.

\subsubsection{Passive components}
Transmission lines, transformers, and loads are modeled using standard dynamic network representations that explicitly account for the electromagnetic energy storage associated with line inductances and shunt capacitances \cite{Kundur1994,Machowski2020}. Transmission lines are described using lumped or distributed-parameter $\pi$-equivalent models, while transformers are represented by dynamic equivalent circuits including leakage inductances, winding resistances, and magnetizing branches \cite{AndersonFouad2003}. Load dynamics are captured using aggregate models that may include voltage- and frequency-dependent behavior, as well as equivalent inductive and capacitive elements \cite{Kundur1994}. 

When combined with \ac{sg} and control models, these component representations give rise to a coupled set of nonlinear differential–algebraic equations describing the overall power system dynamics. Around a given steady-state operating point, the resulting system admits a well-defined linearization, enabling eigenvalue-based small-signal stability analysis while preserving the relevant network dynamic effects at the selected time scale \cite{Machowski2020}.  The details of the adopted models are available in a public repository \cite{code}.

\subsection{Grid-following IBR modeling and control}
\label{ssec:cig}
\begin{figure}[!t]
\centering
\includegraphics[width=0.8\columnwidth]{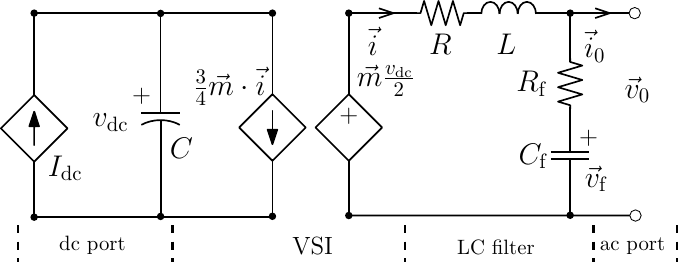}
\caption{Dynamic circuit schematic of the two-level three-phase \ac{ibr} considered in the paper.}
\label{fig:VSI_power}
\end{figure}

\begin{figure}[!t]
\centering
\includegraphics[width=1\columnwidth]{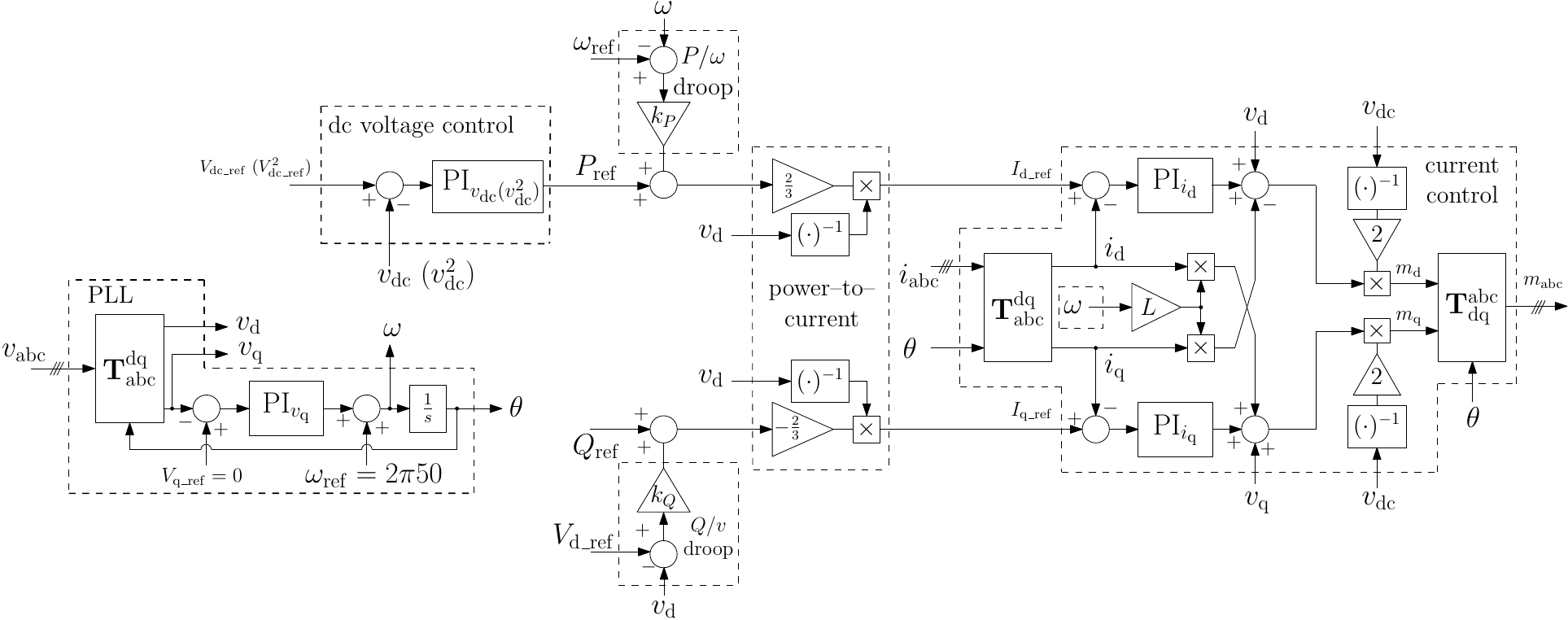}
\caption{Block diagram of a single VSI control scheme. In the figure, $\mathbf{T}_\mathrm{abc}^\mathrm{dq}$ and $\mathbf{T}_\mathrm{dq}^\mathrm{abc}$ are the $\mathrm{abc} \leftrightarrow \mathrm{dq}$ transformation matrices. The symbol $\mathrm{PI}_x$ represents a PI controller ($k_p+\frac{K_i}{s}$) acting on the error $X_\mathrm{ref}-x$.}
\label{fig:VSI_control}
\end{figure}

IBR-based power plants are currently realized using a large number of relatively small two-level three-phase voltage source inverters (VSIs), each including a power source and a dc link at the dc port, and an LC filter at the ac port. The model of a single VSI is illustrated in Fig.~\ref{fig:VSI_power}. In the figure, vectorized signals represent the $dq$ components according to $\vec{x}=x_d+jx_q$, and the operator $\cdot$ denotes the dot product between two vectors. The control scheme, illustrated in Fig.~\ref{fig:VSI_control}, consists of five distinct control stages. The current control, implemented in a synchronous $dq$ frame, decouples the regulation of the $d$- and $q$-axis currents by means of feedforward terms and drives them to their reference values using identical \ac{pi} regulators. The current references are obtained from active and reactive power commands through a power-to-current transformation \cite{yazdani2010}. To enable power sharing and voltage regulation, nominal references are adjusted using $P/\omega$ and $Q/v$ droop control laws. This approach assumes a stiff grid at the \ac{pcc} and enforces a zero $q$-axis voltage component. The latter is achieved through a \ac{pll}, which synchronizes the inverter to the grid by measuring the voltage at the \ac{pcc}, transforming it into the $dq$ frame, and driving $v_q$ to zero via a PI controller \cite{chung2000phase}. Finally, the control scheme includes regulation of the dc-port voltage to a desired reference value. This objective can be achieved either by directly controlling the dc-link voltage, i.e., $v_\mr{dc}$, or by controlling the energy stored in the dc-link capacitor, i.e., $v_\mr{dc}^2$ \cite{yazdani2010}. In both cases, the control action is implemented through a \ac{pi} regulator. 

In summary, the control architecture relies on the tuning of three \ac{pi} controllers, namely those associated with current control, the \ac{pll}, and the dc-port voltage regulation, as well as two droop gains governing the $P/\omega$ and $Q/v$ characteristics.  

In practice, IBR power plants consist of a large number of VSIs connected to a common bus through a collector system. Dynamic aggregation of VSI systems connected to a common busbar has been partially investigated in \cite{Mishra2024}. In this paper, the aggregation procedure is extended to include all IBR physical and control parameters shown in Figs.~\ref{fig:VSI_power} and~\ref{fig:VSI_control}. The resulting aggregated parameters are summarized in Table~\ref{table:aggregation}.

\begin{table}[t]
\centering
\caption{Dynamic aggregation of $N$ VSI devices.}
\begin{tabular}{@{}lc@{}}
\toprule
\textbf{Single VSI} & \textbf{Aggregated value} \\ \midrule

\multicolumn{2}{@{}l}{\textbf{Physical elements}} \\ 
$I_\mathrm{dc}$, $C$, $C_\mathrm{f}$ & $N$ $\times$ \\ 
$R$, $L$, $R_\mathrm{f}$ & $1/N$ $\times$ \\ 
\multicolumn{2}{@{}l}{\textbf{Control Parameters}} \\ 
$PI_{v_\mathrm{dc}(v_\mathrm{dc}^2)}$, $PI_{i_\mathrm{dq}}$ & $N$ $\times$ \\ 
$PI_{v_\mathrm{q}}$ & $1$ $\times$ \\ 
$k_\mathrm{PQ}$ & $N$ $\times$ \\ 

\multicolumn{2}{@{}l}{\textbf{Reference Values}} \\ 
$Q_{\rm ref}$ & $N$ $\times$ \\ 
$V_\mathrm{dc\_ref}~(V_\mathrm{dc\_ref}^2)$, $V_\mathrm{d\_ref}$, $\omega_\mathrm{ref}$ & $1$ $\times$ \\ 

\bottomrule
\end{tabular}
\label{table:aggregation}
\end{table}

\subsection{System level scenario synthesis}
Small-signal stability analysis is inherently local in nature and depends on the steady-state operating point about which the power system dynamics are linearized \cite{Kundur1994,Machowski2020}. Accordingly, the nonlinear differential–algebraic model comprising \acp{sg}, network components, loads, and associated control systems is first initialized at a feasible equilibrium defined by a power-flow solution. This equilibrium determines consistent values of bus voltages, phase angles, and power injections, as well as the corresponding internal states of generators, IBRs, and controllers.

To account for variability in loading conditions, network configurations, and generation dispatch, multiple representative operating points are considered. Each operating point defines a distinct equilibrium of the nonlinear system and gives rise to a corresponding linearized state-space representation. The resulting family of linear models enables a systematic assessment of small-signal stability properties and modal behavior across a range of plausible system conditions, thereby capturing the sensitivity of electromechanical and control-related modes to changes in the operating point.

\section{Small-signal stability analysis tool}
\label{sec:stability_analysis_tool}
In this study, the power system stability is investigated via direct eigenvalue analysis applied to the full grid linearized model. In this section, we describe the adopted linearization procedure, which is similar to the one proposed in \cite{Collados2020,Collados2024}. 

The linearization is performed in the synchronous $dq$ reference frame. Each subsystem, representing a specific network element (generators, lines, loads, etc.), is individually linearized and modeled by a state-space representation after a proper definition of the associated input and output vectors. The overall system model is subsequently constructed by interconnecting the individual linearized subsystems.
Table~\ref{tab:input-output-linearization} summarizes the input and output variables assigned to each system component. Phasors $\vec{v}$ and $\vec{i}$ are the voltage and current at the point of connection with the rest of the network (e.g., the \ac{pcc} for \acp{sg} and \acp{ibr}). Both phasors are referred to the common global reference frame (slack bus).

\begin{table}[t]
\caption{Input-output definition of the subsystems components.}
\label{tab:input-output-linearization}
\begin{center}
\begin{tabular}{lcc}
\hline
\textbf{Component} & \textbf{Inputs}                                & \textbf{Output} \\ \hline
\ac{sg}            & $P_{\rm ref},V_{\rm ref},\vec{i}$              & $\vec{v}$        \\
\ac{ibr}           & $P_{\rm ref}(I_{\rm dc}), V_{\rm ref},\vec{i}$ & $\vec{v}$        \\
Transformer, RL-branch, RL-load        & $\vec{v}$                                      & $\vec{i}$        \\
Shunt capacitor    & $\vec{i}$                                      & $\vec{v}$        \\ \hline
\end{tabular}
\end{center}
\end{table}

The \ac{sg} linear subsystem is derived from the linearization of the Sauer–Pai formulation \cite{Sauer1998}, together with the steam turbine and governor dynamics \cite{mathworks2024steam,governor}, and the IEEE Type 1 excitation system \cite{mathworks2024excitation,avrIEEE}.

The \ac{ibr} linear subsystem is obtained starting from the model introduced in Section~\ref{ssec:cig}. Both \ac{sg} and \ac{ibr} models include the linearization of the rotation between the internal and the common global $dq$ reference frames. Transformers are represented by the standard T-equivalent circuit, consisting of series RL impedances on the primary and secondary sides, and a shunt branch modeling core losses and magnetizing effects through a parallel RL network. Loads are represented by constant RL branches, while transmission lines are modeled through the conventional $\pi$-equivalent circuit composed of a series RL element and two shunt capacitances. Since these elements exhibit linear behavior, no linearization is required for loads, lines, or shunt capacitors.

\begin{figure}[!t]
    \centering
    \includegraphics[width=1\linewidth]{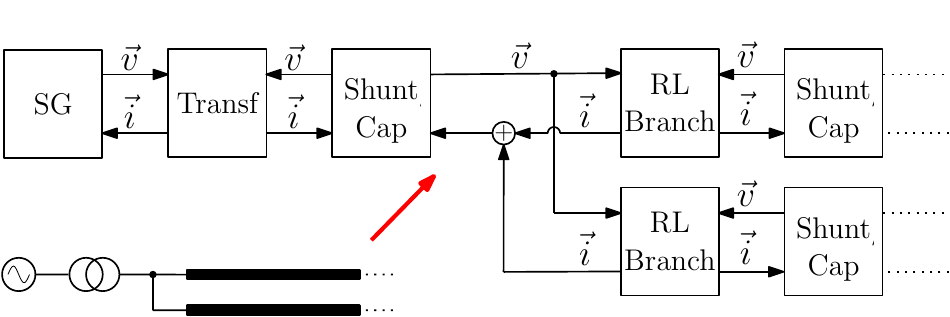}
    \caption{Linearization tool, example of subsystems connection.}
    \label{fig:connection_example}
\end{figure}

To illustrate how the linear subsystems are interconnected, we refer to the example shown in Fig.~\ref{fig:connection_example} where the small portion of the grid at the bottom is modeled through the interconnection of several subsystems. Specifically, the \ac{sg} subsystem receives the current from the transformer and provides the corresponding voltage output; the transformer subsystem, in turn, takes as inputs the voltages from both the \ac{sg} and the shunt capacitor (part of the $\pi$-line model) and outputs the related currents. When multiple branches are involved—such as between the two lines and the transformer in our example—the current interconnections must be established in accordance with the Kirchhoff law.
In mathematical terms, the interconnection of the subsystems is established through a suitable algebraic composition of their state-space matrices, namely $A_i, B_i, C_i,$ and $D_i$, with $i$ denoting each individual subsystem. This procedure yields the overall linear representation of the entire power system. Among the resulting matrices, the global state matrix $A_{ps}$ can subsequently be employed to perform eigenvalue-based stability analyses.


The linearization process requires the steady-state values of the state variables associated with the nonlinear subsystems (i.e., \ac{sg} and \ac{ibr}). These quantities are derived from power flow results.
The developed code and the details of the adopted models are available in a public repository \cite{code}.

\section{Identification of stability manifolds}
\label{sec:stability_manifolds}
The principal aim of this work is to investigate the sensitivity of power system stability to variations in the control parameters of the \acp{ibr}. To this end, we propose an \ac{asm} capable of identifying the boundaries of the manifolds, within a predefined parameter domain, where the power system is stable.
Given a set of $N_s$ network scenarios, the power system is considered stable if small-signal stability holds in all of these scenarios.

Let $\rho \in \mathbb{R}^{N_{\rho}}$ denote the vector composed by the parameters under analysis. The parameter domain is defined by intervals $[\rho_j^{min},\rho_j^{max}]$, $j=1,2,\ldots,N_{\rho}$. We introduce the function \texttt{isPSstable}($\rho$), which assign a stability label $s$ to the given parameter vector $\rho$: $s=1$ if the power system is stable in all the $N_s$ network scenarios considered, and $s=0$ if at least one scenario is unstable. Each evaluation of \texttt{isPSstable}($\rho$) involves computing $N_s$ linearized model of the power system (as detailed in Section~\ref{sec:stability_analysis_tool}) given the parameter values specified by $\rho$. This requires computing $N_s$ state matrices $A_{ps}$ and ensuring that no matrix has eigenvalues with nonnegative real parts.
Given the function \texttt{isPSstable}($\rho$), the stability manifold in the parameter domain could, in principle, be determined by exhaustively scanning all possible parameter combinations with a chosen discretization step. Nevertheless, the computational cost of evaluating \texttt{isPSstable}($\rho$) is high, as each evaluation requires, at minimum, the repeated computation of the eigenvalues of a large-scale system matrix. To face this issue, we propose the \ac{asm} that proceeds as follows:

\begin{enumerate}[leftmargin=*]
    \item Generate an initial random sampling set $G_{init}$ with size $N_{init}$ within the parameter domain, following a uniform distribution.
    \item For each sample $\rho_i$, $i=1,2,\ldots,N_{init}$, evaluate the function \texttt{isPSstable}($\rho_i$) to assign the corresponding stability classification label: stable ($s_i$=1) or unstable ($s_i$=0).
    \item Use the labeled dataset to train a probabilistic classifier---specifically, an \ac{svm} with posterior calibration \cite{Cristianini2000}---which estimates the probability of system stability across the parameter domain. The trained model provides the predicted stability probability for any given parameter vector $\rho$, denoted as $P_r^{SVM,0}(\rho)$.
    \item To improve the accuracy of the stability boundary, generate a larger set of candidate points ($N_r>>N_{init}$) within the parameter domain and evaluate their predicted stability probabilities using the trained \ac{svm}. Select the $N_a$ candidates whose predicted probabilities are closest to a predefined threshold $P_r^{th}$. These samples are added to the initial dataset to form an augmented training set $G_{ai}$ (of size $N_a+N_{init}$), which is then used to retrain the classifier, yielding the final probability function $P_r^{SVM}(\rho)$.
\end{enumerate}

This probabilistic method inherently represents a tradeoff between the accuracy achievable through a deterministic analysis and the computational cost.
\section{Optimal tuning of IBR control parameters}\label{sec:optimal_gains}
Despite the principal aim of this study is stability analysis, we need to define an initial value for the \ac{ibr} control parameters. This is necessary to show the result of the analysis. The number of the considered parameters $N_{\rho}$ is indeed generally higher than two, and visualizing the obtained stability domains is not feasible up to the three-dimensional case. Therefore, we will apply the \ac{asm} to different couples of control parameters while keeping fixed the remaining ones to the initial values. Therefore, in this section, we introduce the method adopted to define such initial values optimally. In our specific case, we consider six parameters, i.e., the PI gains of current control $(k_p^i,k_i^i)$, \ac{pll} $(k_p^{pll},k_i^{pll})$ and dc voltage control, in the two alternative cases of $v_{\rm dc}$ control $(k_p^{dc},k_i^{dc})$ and $v_{\rm dc}^2$ control $(k_p^{2dc},k_i^{2dc})$.

The idea is to consider the scenario where engineers have to tune the control parameters of one generic \ac{ibr} to be connected to the grid. The first step is to define the parameter region that can be explored to get the optimal values. This is realized by adopting as guideline a set of rules mainly based on the relation between the bandwidth of the different control loops. More specifically, we consider the following conditions:
\begin{enumerate}
    \item[C1)] The current control loop must be significantly faster than the \ac{pll}, i.e., $\omega_{c}^i > 10 \omega_c^{pll}$ \cite{Teodorescu:2011,Harnefors:2007};
    \item[C2)] the dc voltage control must be slow enough not to react to the intrinsic $2 \omega^{nom}$ ripple, \textit{i.e},  $\omega_{c}^{dc} < 2 \omega^{nom}$\cite{Yazdani:2010};
    \item[C3)] all control loops must have a phase margin $\phi_m>45^\circ$, as required for robust damping and stability \cite{Teodorescu:2011,Harnefors:2007}.
\end{enumerate}

No specific constraint is imposed between the \ac{pll} and dc voltage control bandwidths, since such separation is not generally required \cite{Yazdani:2010,Cespedes:2014,Harnefors:2015}. Under these conditions, given a value $\rho$ for the considered parameters, it is possible to compute a closed-form expression for control loop bandwidths and phase margins under standard simplifying assumptions widely adopted in the literature \cite{Liserre:2005,Teodorescu:2011,Harnefors:2007,Cespedes:2014} (modeling the grid and the converter as ideal voltage sources; expressing the dynamics in the synchronous $dq$ frame; assuming perfect decoupling of cross-coupling terms; and linearizing around the operating point). Let us indicate this computation as $[\omega_c^i,\omega_c^{pll},\omega_c^{dc},\phi_{m}^i,\phi_{m}^{pll},\phi_{m}^{dc}] = \texttt{BwPm}(\rho)$.

Then, we consider a set of $N_c$ possible connections of the \ac{ibr} to the grid, evaluated over $N_s$ network scenarios. For each candidate combination of \ac{ibr} connections and network scenario, the overall system eigenvalues can be computed using the tool introduced in Section~\ref{sec:stability_analysis_tool}. Based on this, we implement an algorithmic function that returns the maximum real part 
of the overall system eigenvalues obtained across all $N_c$ connection combinations and $N_s$ network scenarios: $\alpha_{max} = \texttt{PSSA}(\rho)$.    

The optimal tuning of the control parameters is finally formalized as follows:
\begin{align}
\rho^* &= \arg\min_{\rho} \alpha_{max} \label{eq:cost_function}\\
\text{s.t.}\quad 
& \alpha_{max} = \texttt{PSSA}(\rho) \label{eq:con1} \\
& \alpha_{max} \leq -\varepsilon \label{eq:con2} \\
& [\omega_c^i,\omega_c^{pll},\omega_c^{dc},\phi_{m}^i,\phi_{m}^{pll},\phi_{m}^{dc}] 
   = \texttt{BwPm}(\rho) \label{eq:con3}\\
& \omega_{c}^i \geq 10 \omega_c^{pll}, \quad 
  \omega_c^{dc} \leq 2\omega^{nom} \label{eq:con5} \\  
& \phi_m^i \geq 45^{\circ}, \quad 
  \phi_m^{pll} > 45^{\circ}, \quad 
  \phi_m^{dc} > 45^{\circ} \label{eq:con6} \\
& \rho_j^{min} \leq \rho_j \leq \rho_j^{max} 
  \quad \forall j = 1,\ldots,N_{\rho} \label{eq:con7}
\end{align}


In constraint \eqref{eq:con2}, $\varepsilon>0$ is arbitrarily small; it is used to impose that $\alpha_{max}$ is strictly lower than zero, meaning that the system is asymptotically stable. The remaining constraint imposes the above-introduced conditions C1)--C3). The optimization problem \eqref{eq:cost_function}–\eqref{eq:con7} is inherently nonlinear and non-analytic. Constraint \eqref{eq:con1} requires computing the system eigenvalues through the numerical routine \texttt{PSSA}, so the resulting dependence on the parameter vector $\rho$ cannot be expressed in closed form. Constraint \eqref{eq:con3}, although analytically expressible, remains nonlinear. Therefore, classical gradient-based or convex optimization methods are not applicable.

To address this issue, a heuristic derivative-free strategy is adopted. In particular, optimization is performed using the surrogate algorithm implemented in the MATLAB function \texttt{surrogateopt}. Surrogate optimization \cite{Jones:1998,matlab_surrogateopt} constructs an approximate model of the objective function over the admissible domain and iteratively refines it by selecting new evaluation points that balance exploration and exploitation. This makes it suitable for expensive, non-smooth, and non-analytic problems such as \eqref{eq:cost_function}–\eqref{eq:con7}.

\section{Case study}
\label{sec:case_study}
\begin{figure}[t]
\centering
\includegraphics[width=1\columnwidth]{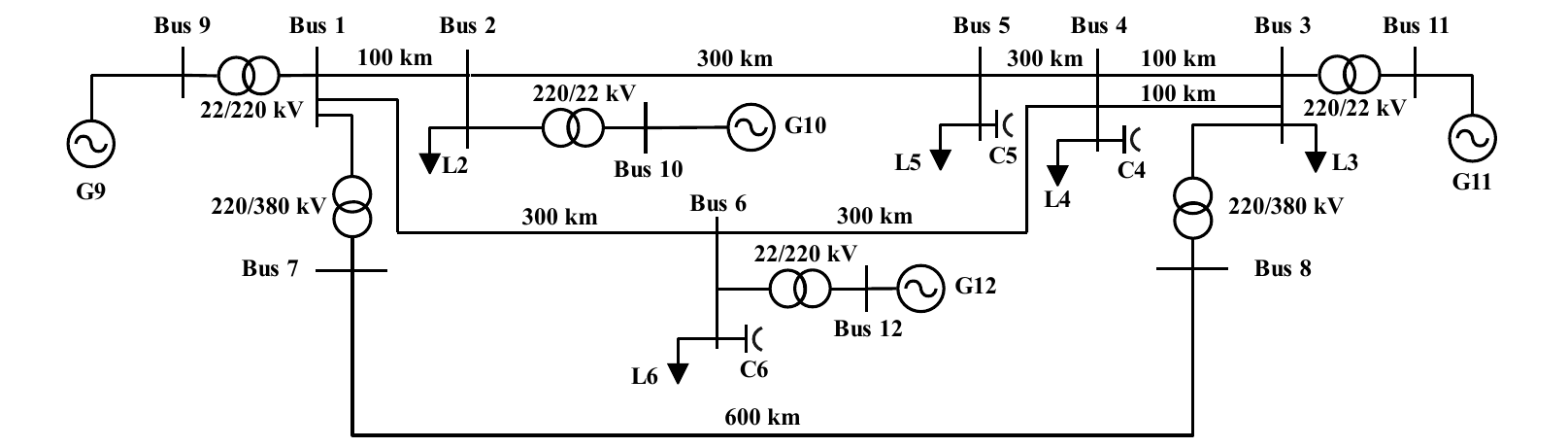}
\caption{Layout of the modified Cigré European HV network.}
\label{fig:cigre}
\end{figure}

The stability analysis approach presented in this study is carried out on the Cigré European HV transmission system described in \cite{cigre}. The network layout is illustrated in Fig.~\ref{fig:cigre}. The grid operates at 50~\texttt{Hz} and includes transmission lines rated at 220~\texttt{kV} and 380~\texttt{kV}. The system consists of 12 buses, with four \acp{sg} connected to the 22~\texttt{kV} buses 9-12. The adopted parameters are mainly those reported in \cite{cigre}, with a number of adjustments and additional specifications detailed below.

Parameter modifications include: assigning a transformer resistance of 2\%, whereas it is assumed zero in the original benchmark;
matching the rated power of the 22/220~\texttt{kV} step-up transformers interfacing the generators to the corresponding generator ratings (the original benchmark specifies higher transformer ratings);
reducing the rated apparent power of generator G10 to 350~\texttt{MVA} (instead of the 700~\texttt{MVA} indicated in the original benchmark model).

Additional modeling details introduced in this work are: generator G9 is represented with the same per-unit parameter set as G11 and assigned a 620~\texttt{MVA} rating, replacing the ideal voltage source representation used in the benchmark model;
all \ac{sg} units are equipped with a steam turbine and governor model as in \cite{mathworks2024steam,governor}, configured with a 5\% droop characteristic, as well as with an IEEE Type-1 \ac{avr} following the implementation in \cite{mathworks2024excitation,avrIEEE}. These control models are not specified in the Cigré report \cite{cigre}.

\begin{figure}[!t]
    \centering
    \includegraphics[width=0.9\columnwidth]{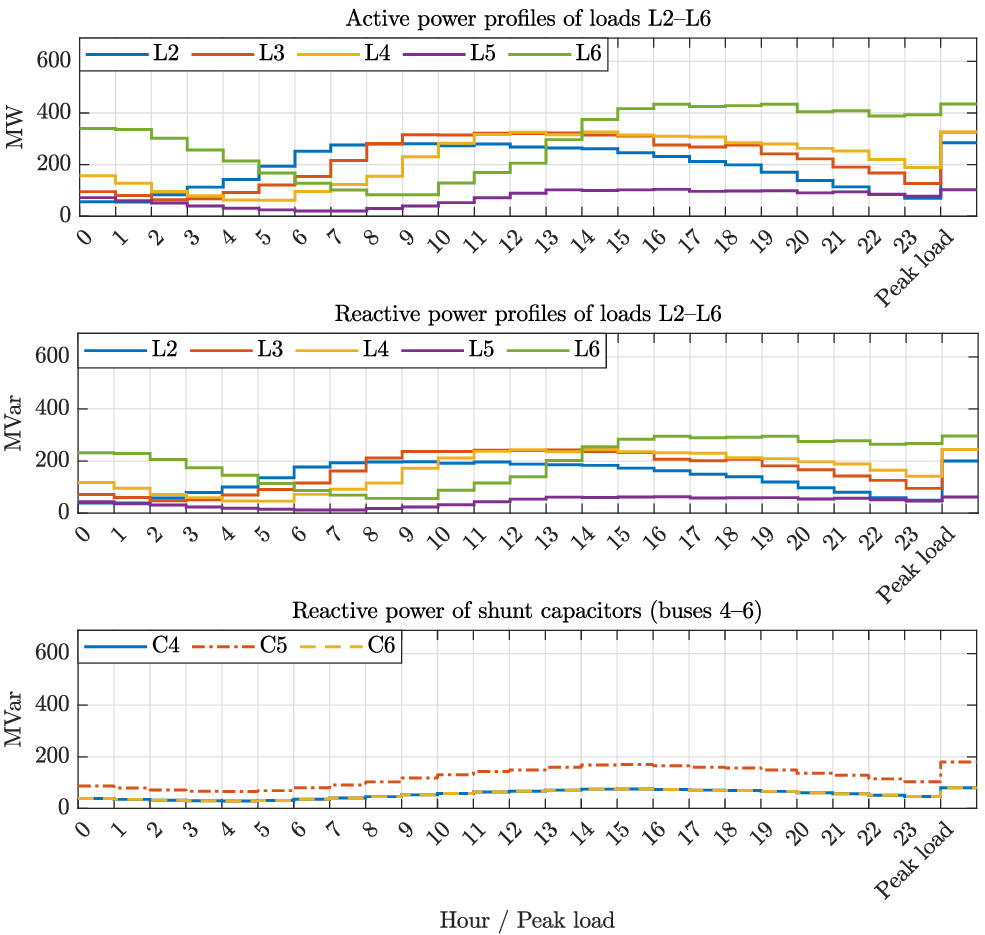}
    \caption{Network scenarios: loads and shunt capacitors profiles.}
    \label{fig:loads_scenarios}
\end{figure}

To study stability under different network configurations, we designed $N_s = 50$ scenarios. Starting from the peak load and shunt capacitor values provided in \cite{cigre}, the 24-hour profiles shown in Fig.~\ref{fig:loads_scenarios} were generated; the figure also reports the corresponding peak values. These profiles were obtained from a common normalized industrial daily curve, which was time-shifted differently for each bus to reflect diverse local demand peaks. A small stochastic perturbation was added to the scaled curves so that each load exhibits realistic variability while preserving the overall Cigré-based shape.

Given the 25 operating points of loads and shunt capacitors in Fig.~\ref{fig:loads_scenarios} (the 24-hour profiles plus the peak value), the 50 scenarios were obtained by dispatching the generator set-points under two different conditions: in Scenarios~1--25, the nominal power of G9 is $S^{\text{nom}}_9 = 620$~\texttt{MVA}; in Scenarios~26--50, it is reduced to $S^{\text{nom}}_9 = 310$~\texttt{MVA}, i.e., halved, to model a loss of inertia and of frequency and voltage regulation capabilities. The assumption is that this \ac{sg} consists of two subunits, and one of them may be out of service, for example, due to maintenance. The overall generator nominal powers are reported in Table~\ref{tab:S_scenarios}, where $S_G^{\text{nom}}$ denotes the total nominal power of the system generators and $S_L^{\text{peak}}$ the peak load apparent power.

\begin{table}[t]
\caption{Generator nominal powers (in \texttt{MVA}).}
\label{tab:S_scenarios}
\begin{center}
\begin{tabular}{ccccccc}
\hline \\
\vspace{-15pt} \\
Scenarios & $S_9^{nom}$ & $S_{10}^{nom}$ & $S_{11}^{nom}$ & $S_{12}^{nom}$ & $S_G^{nom}$ & $S_L^{nom}$ \\
\vspace{-7pt} \\
\hline
\vspace{-5pt} \\
1-25 & 620 & 350 & 500 & 500 & 1970 & 1600 \\
26-50 & 310 & 350 & 500 & 500 & 1660 & 1600 \\ \hline
\end{tabular}
\end{center}
\vspace{-5pt}
\end{table}

\begin{figure}[!t]
    \centering
    \includegraphics[width=0.9\columnwidth]{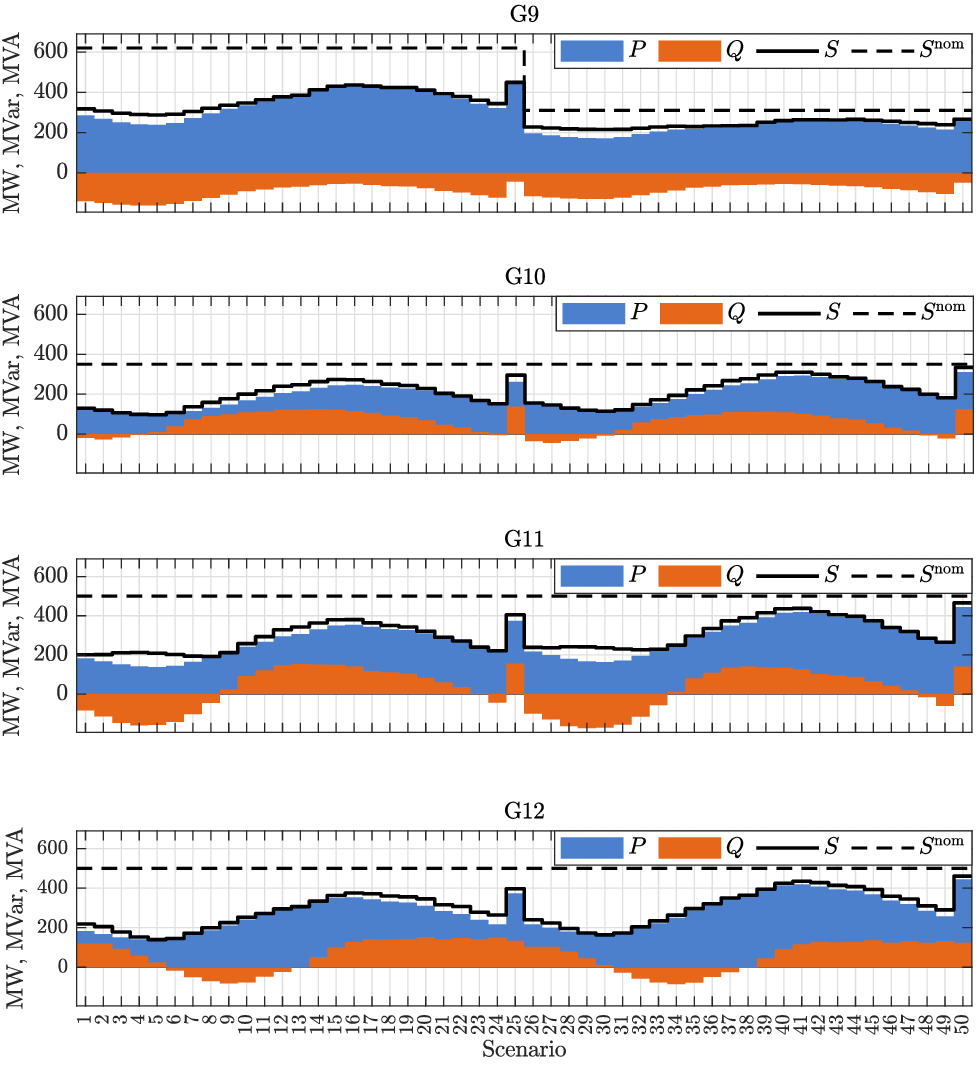}
    \caption{Network scenarios: generator set-points.}
    \label{fig:generators_scenarios}
\end{figure}

The generator set-points reported in Fig.~\ref{fig:generators_scenarios} were computed through a standard Newton-Raphson load-flow calculation. Generators G10--G12 were modeled as PV buses with voltage reference $V_{\text{ref}} = 1.03$~\texttt{pu} and active-power set-points assigned proportionally to the ratios $S^{\text{nom}}_i / S_G^{\text{nom}}$, whereas generator G9 was modeled as the slack bus, also with $V_{\text{ref}} = 1.03$~\texttt{pu}.

It is worth remarking that scenarios were generated with the aim of covering as many plausible operating conditions as possible. In a real-world study, the analysis could rely on actual statistical data for load profiles and on alternative dispatching schemes influenced by market dynamics, grid codes, or operator-specific strategies.

\subsection{IBRs connections}
To show the potentialities of the proposed approach, we analyize the Cigré system considering a progressive replacement of generators G10--G12 with equivalent \acp{ibr} having the same nominal power. Each \ac{sg} is replaced by an aggregated model composed of $N$ identical single IBRs, according to the rules reported in Table~\ref{table:aggregation}.

As the reference single \ac{ibr} unit, the largest commercially available inverter was selected, which, to the best of the authors' knowledge, is manufactured by Gamesa Electric \cite{gamesa}. This inverter is rated at 5~\texttt{MVA}, with a dc-side voltage of 1500~\texttt{V}, and synthesizes an ac output of 660~\texttt{V} at 50~\texttt{Hz}. Accordingly, the physical components of the single \ac{ibr} model have been sized as reported in Table~\ref{tab:ibr_parameters}. Referring to the nominal powers reported in Table~\ref{tab:S_scenarios}, the replacement of G10 is obtained with $N=70$, the one of G11 and G12 with $N=100$.  

\begin{table}[!t]
\begin{center}
\caption{Single base IBR physical parameters definition\\Base values:  5~ \texttt{MVA} / 660 \texttt{V} / 50 \texttt{Hz},   1500 \texttt{V} dc.} 
\label{tab:ibr_parameters}
\begin{tabular}{lcc c}
\hline
\textbf{Description}                 & \textbf{Symbol} & \textbf{Value} & Unit \\ \hline
Filter resistance                    & $R$             &   0.05 & \texttt{pu}        \\
Filter inductance                    & $L$             &   0.15 & \texttt{pu}       \\
Filter capacitance                   & $C_{\rm f}$           &   0.05 & \texttt{pu}       \\
Filter shunt resistance              & $R_{\rm f}$           &   0.0016 & \texttt{pu}      \\ 
dc port capacitance                  & $C$             &   15 & \texttt{mF}           \\ \hline
\end{tabular}
\end{center}
\vspace{-5pt}
\end{table}

Finally, the potential \acp{ibr} connection combinations considered for the stability analysis are $N_c=7$ accounting for the replacement of one generator among G10--G12 (three cases), two generators (three cases), and all three generators (one case).

\subsection{Initial tuning of IBR parameters and definition of regions of practical interest} \label{ssec:tuning}
The analysis proposed in this study is focused on the sensitivity of the system stability to variations in the control parameters of the \acp{ibr}. As mentioned in Section~\ref{sec:optimal_gains}, the considered parameters are the PI gains of current control $(k_p^i,k_i^i)$, \ac{pll} $(k_p^{pll},k_i^{pll})$ and dc voltage control, in the two alternative cases of $v_{\rm dc}$ control $(k_p^{dc},k_i^{dc})$ and $v_{\rm dc}^2$ control $(k_p^{2dc},k_i^{2dc})$. Frequency and voltage droop gains are instead always fixed to $k_P = 1/0.05$ \texttt{pu} and $k_Q = 1/1.1$.  

\begin{figure}[!t]
    \centering
    \includegraphics[width=0.9\columnwidth]{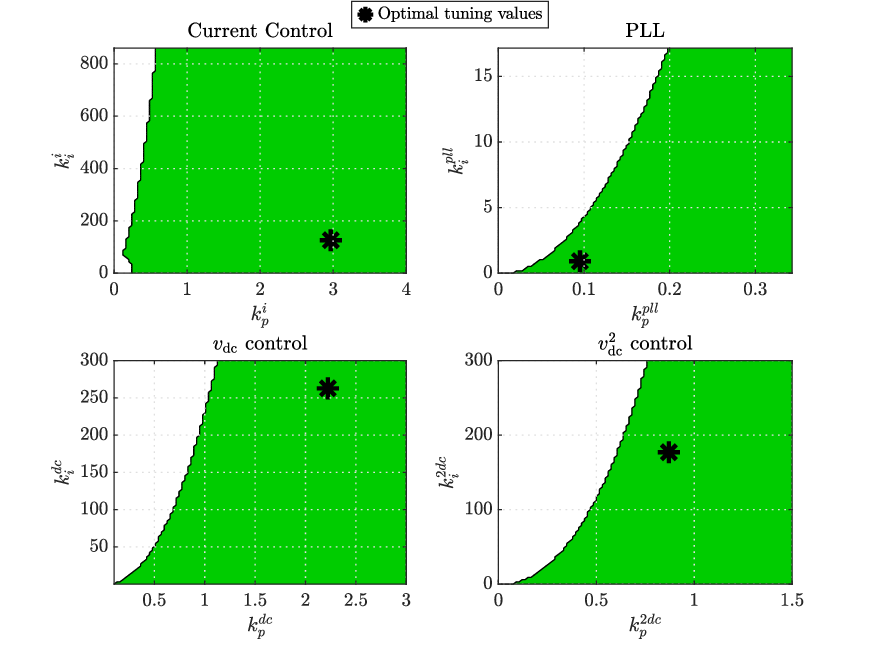}
    \caption{Control parameters regions of practical interest (RPIs) (green) and initial optimal tuning values.}
    \label{fig:initial_tuning}
    \vspace{-8pt}
\end{figure}

\begin{figure*}[!t]
    \centering
    \includegraphics[height=0.25\textheight,keepaspectratio,trim=2.3cm 0 2cm 0,clip]{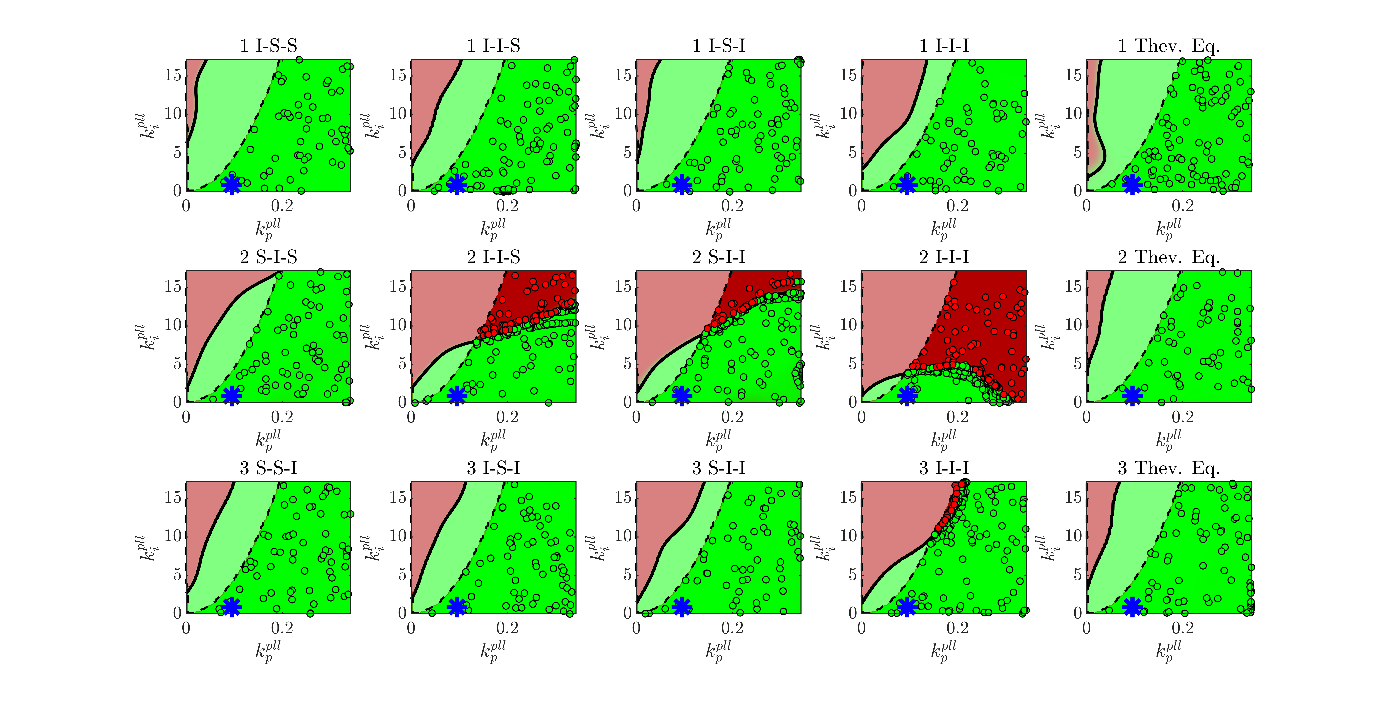}%
    \hspace{5mm}%
    \includegraphics[height=0.24\textheight,keepaspectratio,trim=0 0 0 0.3cm,clip]{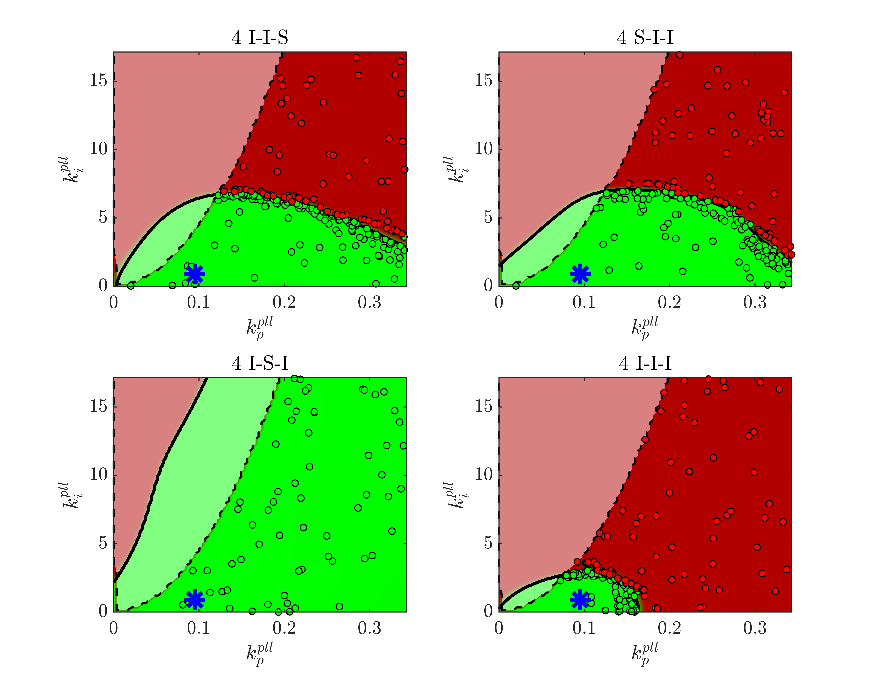}
    \vspace{-5pt}
    \caption{\ac{asm}-based analysis over the \ac{pll} gains, \ac{ibr} version with $v_{\rm dc}$ control. Left plots: case study sets 1-3; right plots: case study set 4. Green areas: stability manifolds; red areas: instability manifolds; dashed line: boundary of \acp{rpi}; blue star: optimal tuning values.}
    \label{fig:pll_ver1}
    \vspace{-5pt}
\end{figure*}

\begin{figure}[!t]
    \centering
    \includegraphics[width=0.9\columnwidth]{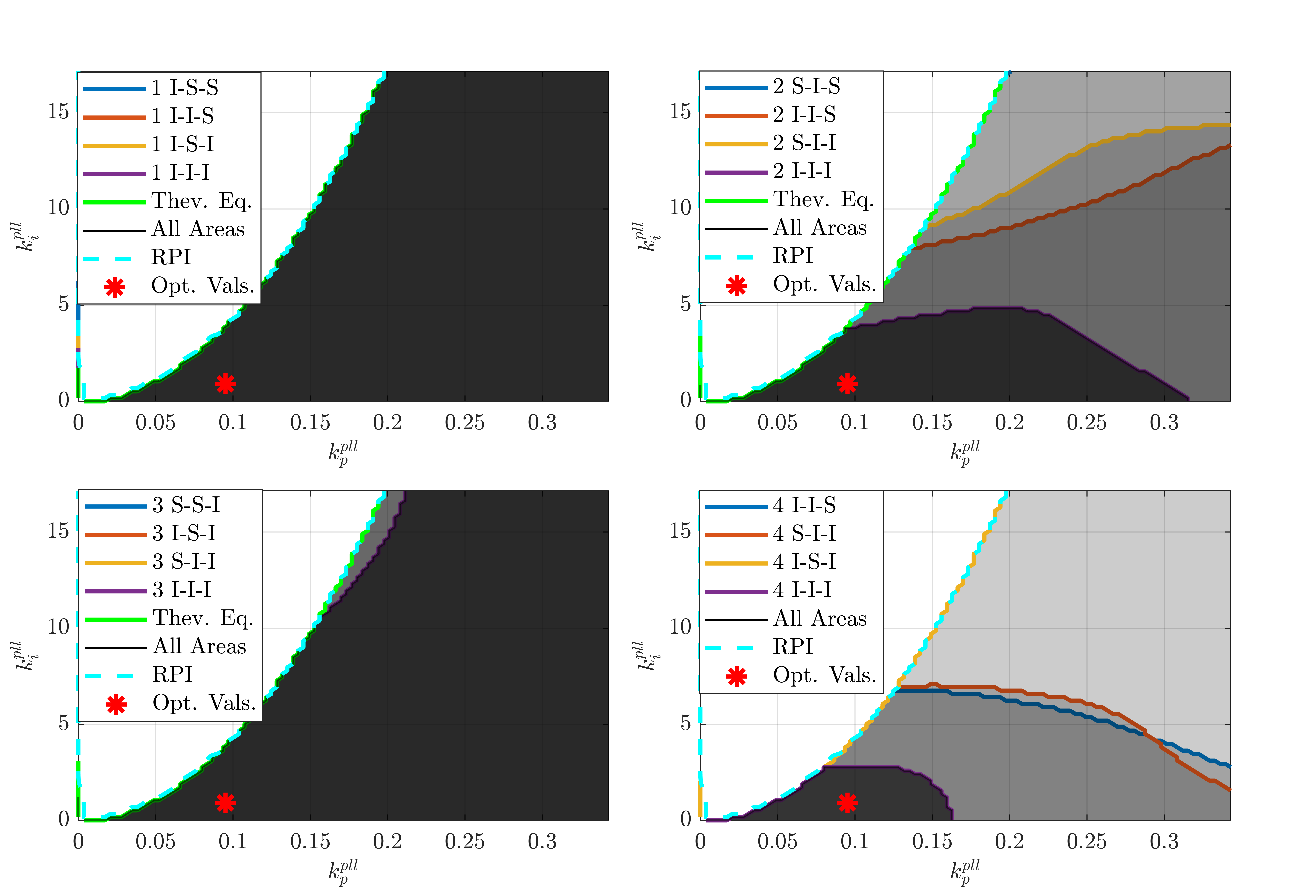}
    \caption{Stability manifolds for \ac{pll} gains, \ac{ibr} with $v_{\rm dc}$ control.}
    \label{fig:pll_ver1_summary}
\end{figure}

\begin{figure}[!t]
    \centering
    \includegraphics[width=0.9\columnwidth]{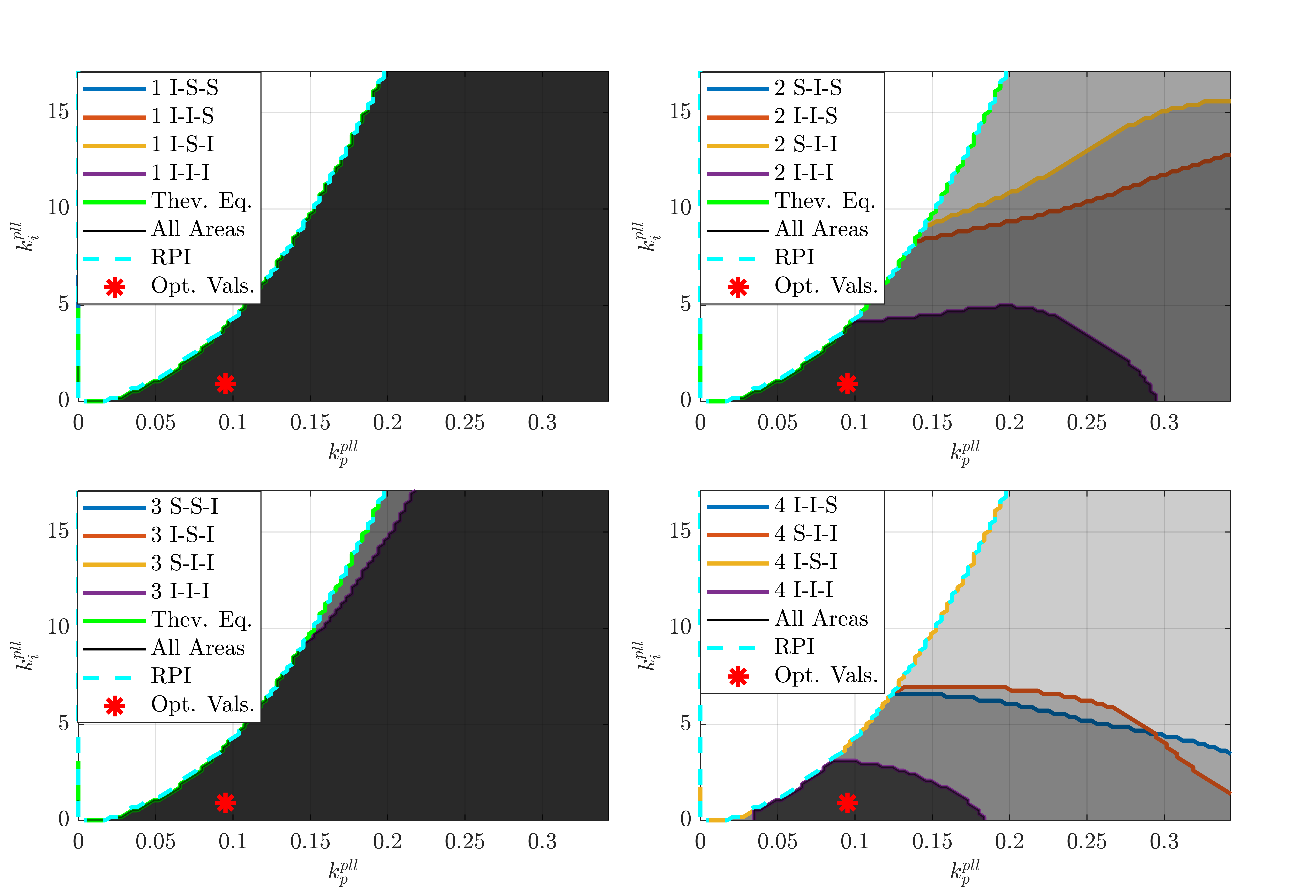}
    \caption{Stability manifolds for \ac{pll} gains, \ac{ibr} with $v_{\rm dc}^2$ control.}
    \label{fig:pll_ver2_summary}
\end{figure}

Table~\ref{tab:ibr_control_parameters} reports the parameters domains used in the \ac{asm} (i.e., $[\rho_j^{min},\rho_j^{max}]$). Given these domains, the optimal tuning algorithm introduced in Section~\ref{sec:optimal_gains} has been applied considering the $N_s=50$ scenarios and $N_c=7$ \acp{ibr} candidate connections above described. Fig.~\ref{fig:initial_tuning} shows the obtained optimal tuning values. We underline that from Fig.~\ref{fig:initial_tuning} to Fig.~\ref{fig:vc_ver2_summary} the values of the parameters are reported in per unit. 

In the same figure, we report the areas obtained for each pair of PI gains by imposing conditions \eqref{eq:con5}--\eqref{eq:con7}, while the remaining control parameters are fixed at their optimal values, i.e., as defined by the algorithm described in Section~\ref{sec:optimal_gains}. Since constraints \eqref{eq:con5}--\eqref{eq:con7} are defined by adopting a set of well-established control engineering rules regarding dynamical performance and internal \ac{ibr} stability margins, we refer to these areas as \ac{rpi}. 

Indeed, as already remarked in Section \ref{sec:optimal_gains}, the general idea of this paper is to consider a scenario in which engineers are required to tune the control parameters of the \ac{ibr} and investigate the resulting effects on the overall system stability by means of the \ac{asm} introduced in Section \ref{sec:stability_analysis_tool}. The final outcome of the \ac{asm} consists of stability manifolds identified within the original parameter domains. 

Being purely based on small-signal stability analysis, these manifolds may include points lying outside the \acp{rpi}. Nevertheless, such results have limited practical relevance, since these operating conditions would not be considered in practice.
Therefore, in the results presented in the following, the stability manifolds are obtained by excluding the points lying outside the \acp{rpi}.

\begin{table}[!t]
\begin{center}
\caption{IBR control parameter domains\\Base values:  5~ \texttt{MVA} / 660 \texttt{V} / 50 \texttt{Hz},   1500 \texttt{V} dc.} 
\label{tab:ibr_control_parameters}
\begin{tabular}{lll c}
\hline
\textbf{Description}                 & \textbf{Symbol} & \textbf{Value} & Unit \\ \hline
PLL control proportional gain        & $k_p^{pll}$     &  [0 \ 0.35] & \texttt{pu}        \\
PLL control integral gain            & $k_i^{pll}$     &  [0 \ 17] & \texttt{pu}       \\
Current control proportional gain    & $k_p^{i}$       &  [0 \ 4] & \texttt{pu}      \\
Current control integral gain        & $k_i^{i}$       &  [0 \ 860] & \texttt{pu}      \\
$v_\mr{dc}$ control proportional gain   & $k_p^{dc}$      &  [0 \ 3] & \texttt{pu}          \\
$v_\mr{dc}$ control integral gain       & $k_i^{dc}$      &  [0 \ 300] & \texttt{pu}      \\
$v_\mr{dc}^2$ control proportional gain & $k_p^{2dc}$     &  [0 \ 1.5] & \texttt{pu}               \\
$v_\mr{dc}^2$ control integral gain     & $k_i^{2dc}$     &  [0 \ 300] & \texttt{pu}             \\ \hline
\end{tabular}
\end{center}
\end{table}

\subsection{Results}
Table \ref{tab:case_studies} introduces all the analyzed case studies. For each one, the table reports the devices connected at buses 10–12 (\ac{sg} or equivalent \ac{ibr}) and the \textit{focus} \acp{ibr} (highlighted by a star and a green background), i.e., \acp{ibr} whose control parameters are varied in the stability analysis. 

Four sets of case studies are identified. In the first three sets, a single focus \ac{ibr} is considered, representing the installation of one new \ac{ibr} in the grid while the remainder of the system is assumed to be fixed. Accordingly, when present, the other \acp{ibr} (highlighted by a yellow background) operate with fixed, initially optimized control parameters. In the fourth case study set, the control parameters of all connected \acp{ibr} are varied simultaneously throughout the analysis.

\begin{table}[!t]
\caption{Case Studies}
\begin{center}
\label{tab:case_studies}
\begin{tabular}{|l|c|c|c|}
\hline
\rowcolor[HTML]{EFEFEF} 
\textbf{Case Study}                      & \textbf{Bus 10}              & \textbf{Bus 11}              & \textbf{Bus 12}              \\ \hline
\rowcolor[HTML]{FFCCC9} 
\cellcolor[HTML]{EFEFEF}{1 I-S-S} & \cellcolor[HTML]{34FF34}IBR$^\star$ & SG                           & SG                           \\ \hline
\cellcolor[HTML]{EFEFEF}{1 I-I-S} & \cellcolor[HTML]{34FF34}IBR$^\star$ & \cellcolor[HTML]{FFFE65}IBR  & \cellcolor[HTML]{FFCCC9}SG   \\ \hline
\cellcolor[HTML]{EFEFEF}{1 I-S-I} & \cellcolor[HTML]{34FF34}IBR$^\star$ & \cellcolor[HTML]{FFCCC9}SG   & \cellcolor[HTML]{FCFF2F}IBR  \\ \hline
\rowcolor[HTML]{FFFE65} 
\cellcolor[HTML]{EFEFEF}{1 I-I-I} & \cellcolor[HTML]{34FF34}IBR$^\star$ & IBR                          & IBR                          \\ \hline
\rowcolor[HTML]{FFCCC9} 
\cellcolor[HTML]{EFEFEF}{2 S-I-S} & SG                           & \cellcolor[HTML]{34FF34}IBR$^\star$ & SG                           \\ \hline
\cellcolor[HTML]{EFEFEF}{2 I-I-S} & \cellcolor[HTML]{FFFE65}IBR  & \cellcolor[HTML]{34FF34}IBR$^\star$ & \cellcolor[HTML]{FFCCC9}SG   \\ \hline
\cellcolor[HTML]{EFEFEF}{2 S-I-I} & \cellcolor[HTML]{FFCCC9}SG   & \cellcolor[HTML]{34FF34}IBR$^\star$ & \cellcolor[HTML]{F8FF00}IBR  \\ \hline
\rowcolor[HTML]{F8FF00} 
\cellcolor[HTML]{EFEFEF}{2 I-I-I} & IBR                          & \cellcolor[HTML]{34FF34}IBR$^\star$ & IBR                          \\ \hline
\rowcolor[HTML]{FFCCC9} 
\cellcolor[HTML]{EFEFEF}{3 S-S-I} & SG                           & SG                           & \cellcolor[HTML]{34FF34}IBR$^\star$ \\ \hline
\cellcolor[HTML]{EFEFEF}{3 I-S-I} & \cellcolor[HTML]{FCFF2F}IBR  & \cellcolor[HTML]{FFCCC9}SG   & \cellcolor[HTML]{34FF34}IBR$^\star$ \\ \hline
\cellcolor[HTML]{EFEFEF}{3 S-I-I} & \cellcolor[HTML]{FFCCC9}SG   & \cellcolor[HTML]{FFFE65}IBR  & \cellcolor[HTML]{34FF34}IBR$^\star$ \\ \hline
\rowcolor[HTML]{FFFE65} 
\cellcolor[HTML]{EFEFEF}{3 I-I-I} & IBR                          & IBR                          & \cellcolor[HTML]{34FF34}IBR$^\star$ \\ \hline
\rowcolor[HTML]{34FF34} 
\cellcolor[HTML]{EFEFEF}{4 I-I-S} & IBR$^\star$                         & IBR$^\star$                         & \cellcolor[HTML]{FFCCC9}SG   \\ \hline
\rowcolor[HTML]{34FF34} 
\cellcolor[HTML]{EFEFEF}{4 S-I-I} & \cellcolor[HTML]{FFCCC9}SG   & IBR$^\star$                         & IBR$^\star$                         \\ \hline
\rowcolor[HTML]{34FF34} 
\cellcolor[HTML]{EFEFEF}{4 I-S-I} & IBR$^\star$                         & \cellcolor[HTML]{FFCCC9}SG   & IBR$^\star$                         \\ \hline
\rowcolor[HTML]{34FF34} 
\cellcolor[HTML]{EFEFEF}{4 I-I-I} & IBR$^\star$                         & IBR$^\star$                         & IBR$^\star$                         \\ \hline
\end{tabular}
\end{center}
\vspace{-10pt}
\end{table}

The stability analysis based on the \ac{asm} has been applied to all the mentioned case studies, over the couples of PI gains for \ac{pll}, current control, and dc voltage control. Both alternative versions of the \ac{ibr} control schemes, namely with $v_{\rm dc}$ control and $v^2_{\rm dc}$ control, have been analyzed. In all cases, the adopted \ac{asm} parameters are: $P_r^{th}=0.8$, $N_{init}=100$, and $N_a=250$.

\subsubsection{PLL gains}
Fig. \ref{fig:pll_ver1} shows the results of the analysis for the \ac{pll} PI control gains with $v_{\rm dc}$ control. Each subplot refers to one of the case studies defined in Table~\ref{tab:case_studies} and reports the estimated probability of stability over the considered parameter space. Full green indicates a probability equal to one, whereas full red corresponds to zero probability. The circles represent the actual sampled points: green circles indicate that the system was found to be stable in all $N_s = 50$ network scenarios (\texttt{isPSstable}($\rho$) = 1), while red circles indicate that instability occurred in at least one scenario (\texttt{isPSstable}($\rho$) = 0). In the figures, we can observe as the \ac{asm} increases the sampling density around the stability boundary, corresponding to the threshold probability $P_r^{th}=0.8$, in order to improve estimation accuracy. As discussed in Section~\ref{ssec:tuning}, regions outside the \ac{rpi} (see Fig.~\ref{fig:initial_tuning}) are excluded by the stability manifolds. This is clearly visible in the plots reported in Fig.~\ref{fig:pll_ver1}. We recall that the \acp{rpi} are defined by conditions C1)--C3), which are well-established control engineering rules regarding dynamical performance and internal \ac{ibr} stability margins. For case study sets 1–3, where a single focus \ac{ibr} is considered, the analysis was also carried out by modeling the network using a Thévenin equivalent, as done in several papers (e.g., \cite{Collados2020,Mohammed2024}), in order to compare the obtained results. This type of analysis is not performed for case study set 4, where more than one focus \ac{ibr} is present; in this case, the network cannot be properly represented by a Thévenin equivalent, which must be defined with respect to a single point of connection. Fig.~\ref{fig:pll_ver1_summary} collects the manifolds obtained for each case study set into a single plot, allowing for a concise comparison. In case study sets 1 and 3 (with the focus \ac{ibr} at buses 10 and 12, respectively), we observe that the actual stability manifolds (green areas) are larger than the \acp{rpi}. Moreover, Fig.~\ref{fig:pll_ver1_summary} shows that, for these case studies, the final stability manifolds are generally coincident. This indicates that, for case study sets 1 and 3, (a) conditions C1)–C3) are effective in ensuring stability, and the final stability manifolds actually coincide with the \acp{rpi}; and (b) the analysis based on the Thévenin equivalent yields results that are consistent with those obtained using the full network model. Concerning point (b), Fig.~\ref{fig:pll_ver1} indicates that the analysis based on the Thévenin equivalent is conservative in these cases, as the actual stability manifolds (i.e., obtained by neglecting the \acp{rpi}) are larger than those derived from the full network model. Such conclusions, however, are not confirmed for case study set 2, which focuses on the \ac{ibr} connected at bus 11. Indeed, with the exception of case 2~S-I-S, the stability manifolds are smaller than the \ac{rpi}. In particular, in case 2~I-I-I (with all three \acp{ibr} are installed) the stability manifold is significantly reduced. This indicates that, in scenarios with high penetration of \acp{ibr}, the tuning of the \ac{pll} gains may become critical. Interestingly, this aspect is not highlighted by the analysis based on the Thévenin equivalent, which returns a stability manifold similar to those obtained in case study sets 1 and 3. Let us now analyze the results obtained for case study set 4, in which the parameters of all the \acp{ibr} present in the system are varied in the analysis. In this case, we observe that, with the exception of case 4~I-S-I, the stability manifolds are smaller than the \ac{rpi}. As already noted for case study set 3, the manifold is particularly reduced in case 4~I-I-I, where all three \acp{ibr} are installed. In this scenario, the reduction is even more pronounced, further highlighting the criticality of \ac{pll} tuning when similar tuning strategies are adopted for all the \acp{ibr} in the system. The fact that case study 4~I-S-I is the only configuration exhibiting a stability manifold coincident with the \ac{rpi} suggests that the presence of an \ac{ibr} at bus 11 is particularly critical for the network. Indeed, this is the only scenario in which an \ac{sg} is connected at that bus. Fig.~\ref{fig:pll_ver2_summary} shows the results obtained for the \ac{pll} gains using the \ac{ibr} version with $v_{\rm dc}^2$ control. Due to space limitations, detailed results such as those reported in Fig.~\ref{fig:pll_ver1} cannot be shown for all the analyzed parameters. Comparing these results with those obtained using the $v_{\rm dc}$ control strategy, shown in Fig.~\ref{fig:pll_ver1_summary}, no significant differences are observed. This suggests that the tuning of the \ac{pll} is not particularly sensitive to the choice between the two dc-bus control methodologies.

\subsubsection{Current control gains}
\begin{figure}[!t]
    \centering
    \includegraphics[width=0.9\columnwidth]{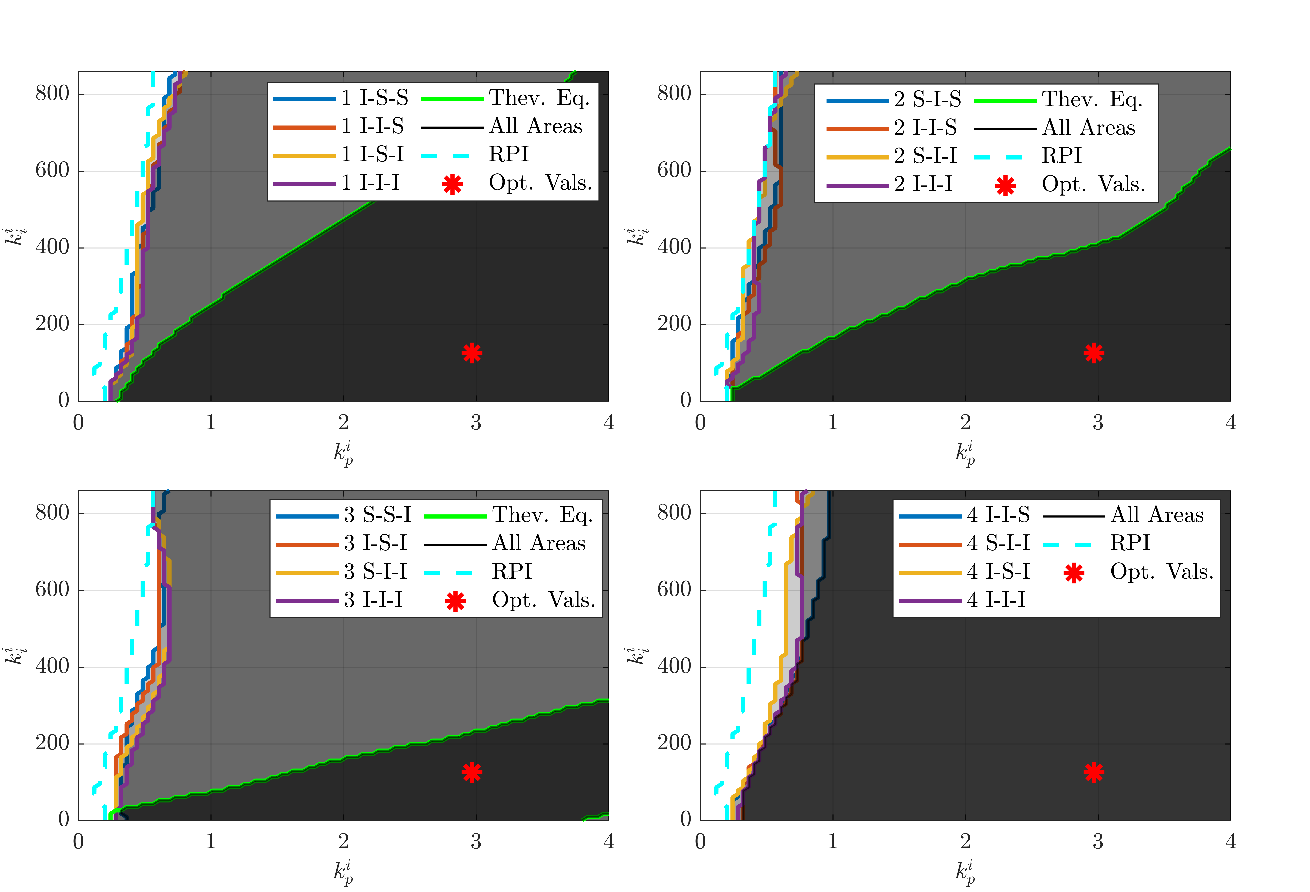}
    \caption{Stability manifolds for current control gains, \ac{ibr} with $v_{\rm dc}$ control.}
    \label{fig:cc_ver1_summary}
\end{figure}

\begin{figure}[!t]
    \centering
    \includegraphics[width=0.9\columnwidth]{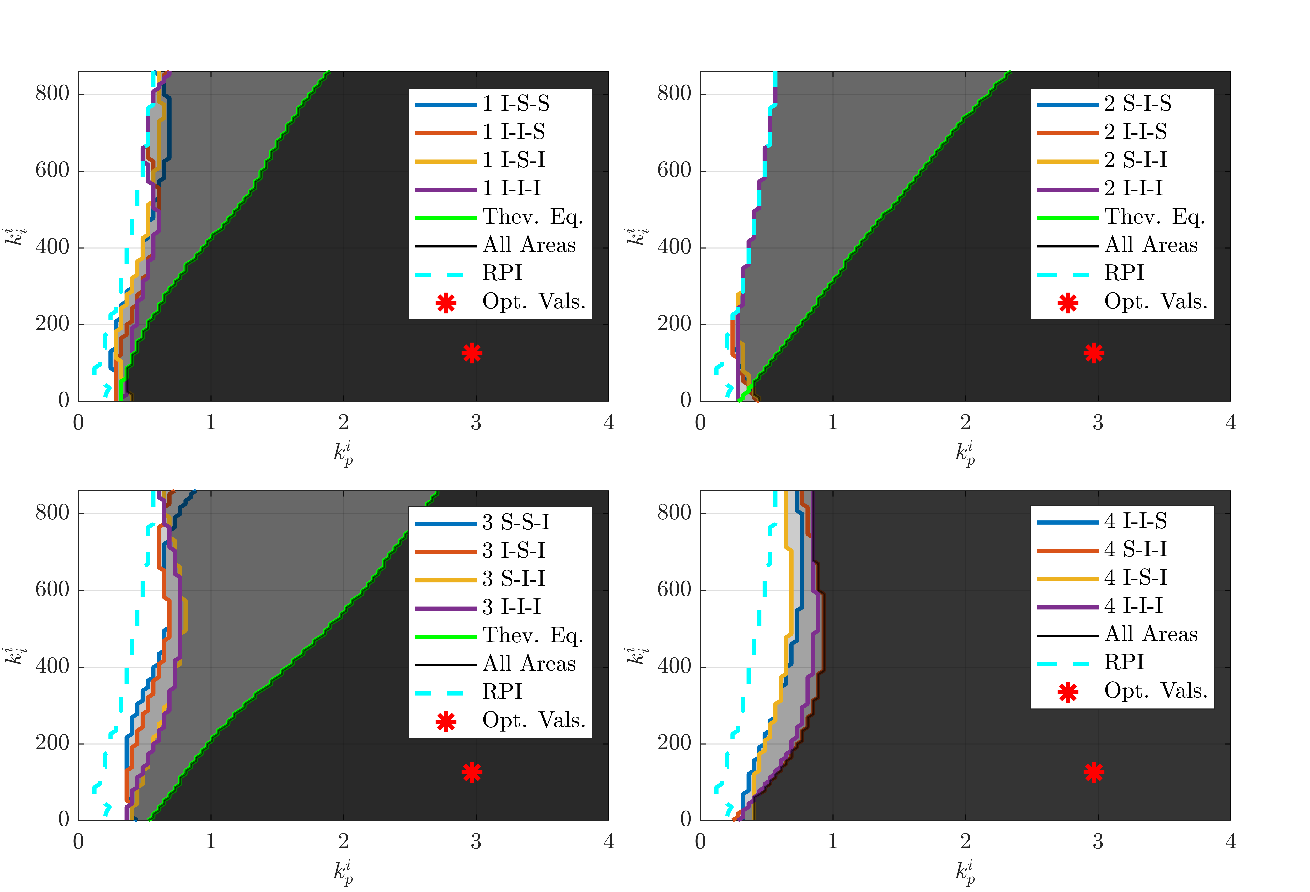}
    \caption{Stability manifolds for current control gains, \ac{ibr} with $v_{\rm dc}^2$ control.}
    \label{fig:cc_ver2_summary}
\end{figure}

Figs.~\ref{fig:cc_ver1_summary} and~\ref{fig:cc_ver2_summary} show the stability manifolds obtained for the current control gains using the two versions of dc voltage control, respectively. As in the case of the \ac{pll} gains, no significant differences are observed between the two control strategies. In addition, the results obtained across the different case study sets are fairly similar. In particular, a close coincidence with the \ac{rpi} is observed, especially in case study sets 1 and 2. In case study sets 3 and 4, the stability manifolds are instead slightly smaller than the \ac{rpi}. This suggests that a minimum proportional gain value higher than that indicated by conditions C1)–C3) should be adopted.
It is interesting to note that, for these parameters, the Thévenin-equivalent-based analysis is overly conservative.

\subsubsection{DC voltage control gains}
\begin{figure}[!t]
    \centering
    \includegraphics[width=0.9\columnwidth]{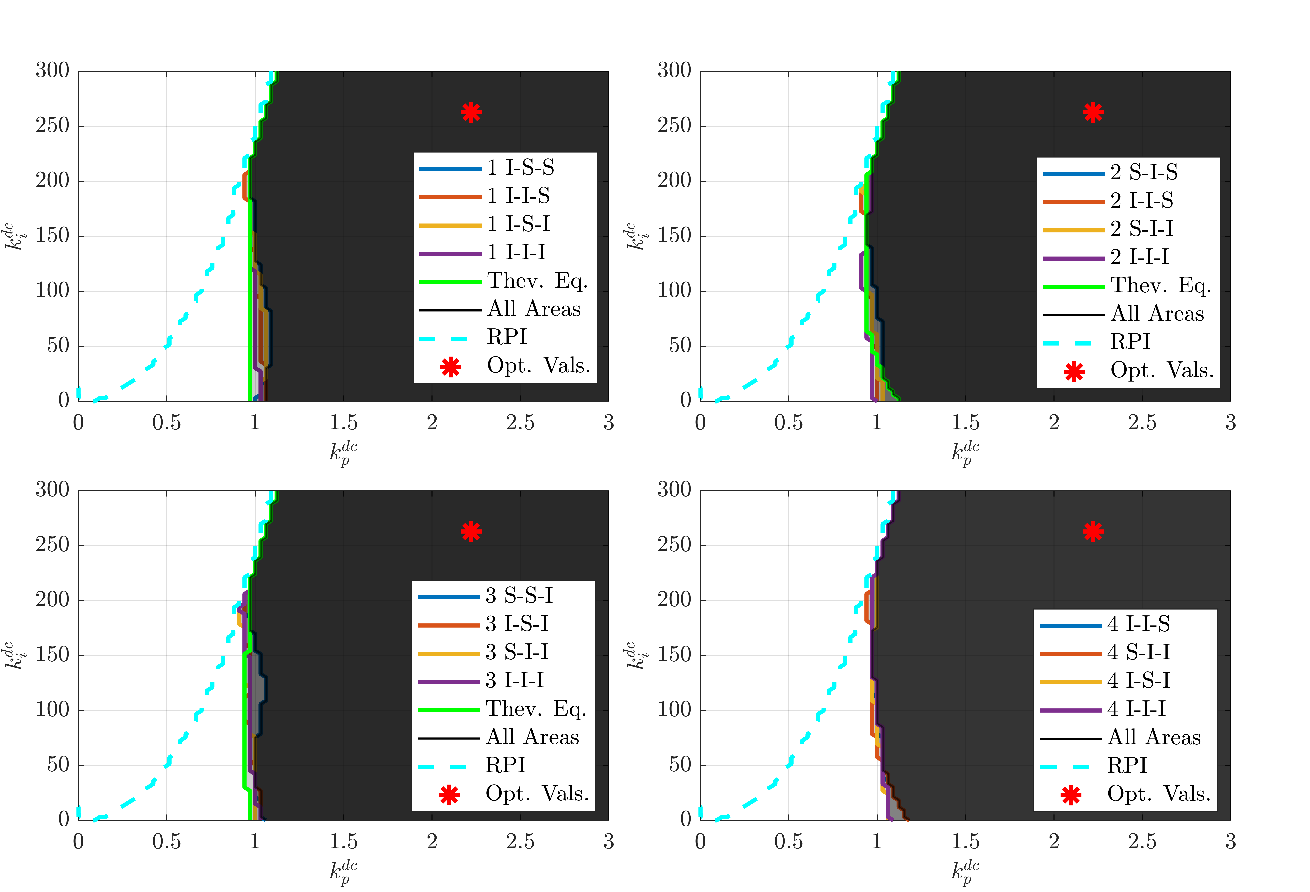}
    \caption{Stability manifolds for $v_{\rm dc}$ control gains.}
    \label{fig:vc_ver1_summary}
\end{figure}

\begin{figure}[!t]
    \centering
    \includegraphics[width=0.9\columnwidth]{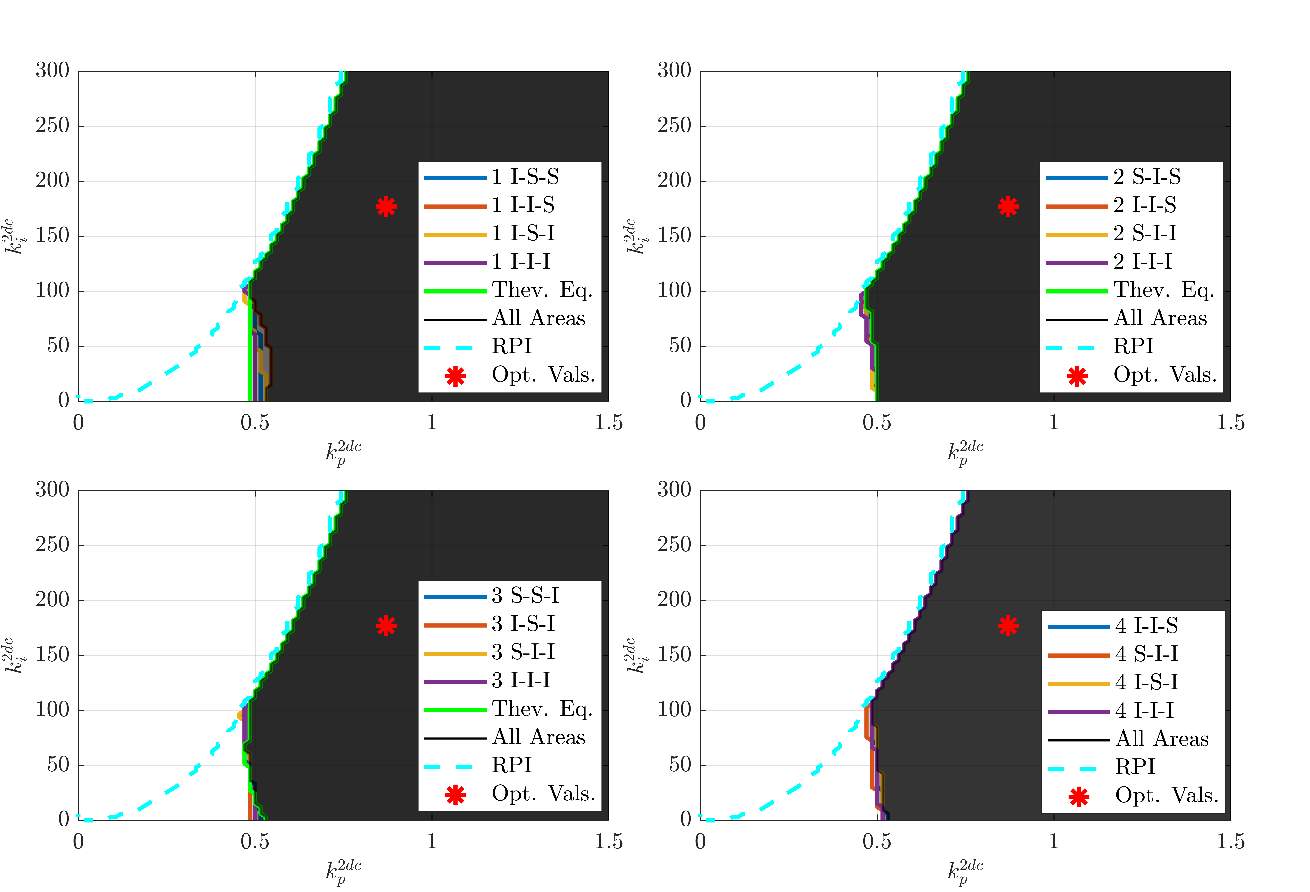}
    \caption{Stability manifolds for $v_{\rm dc}^2$ control gains.}
    \label{fig:vc_ver2_summary}
\end{figure}

Fig.~\ref{fig:vc_ver1_summary} show the results obtained for the $v_{\rm dc}$ control. In this case, no significant differences are appreciated among the case studies. In all cases, the analysis return a minimum value for the proportional gain $k_p^{dc}$ around 1~\texttt{pu} that makes the stability manifold smaller than the \ac{rpi}. Similar results are obtained for the  $v_{\rm dc}^2$ control as depicted in Fig.~\ref{fig:vc_ver2_summary}. In this case the minimum value of the proportional gain $k_p^{2dc}$ is around 0.51~\texttt{pu}.    

\section{Conclusions}
\label{sec:Conclusions} 
This paper introduced the concept of \emph{stability manifolds} and a practical methodology for their identification in power systems with high penetration of \acp{ibr}. The approach combines full-network linearization, multi-scenario eigenvalue analysis, and an \ac{asm} to efficiently map regions of the controller-parameter space that guarantee small-signal stability. Results on a modified Cigré HV benchmark network demonstrate that sensitivity to controller tuning increases with converter penetration. By explicitly accounting for multiple operating scenarios and network configurations, the proposed framework provides a systematic and practically oriented tool to support \acp{tso} and converter designers in controller tuning, interconnection studies, and the definition of admissible parameter ranges.

Future work will extend the proposed framework to incorporate a broader range of converter-interfaced technologies beyond grid-following \acp{ibr}. In particular, the inclusion of grid-forming controls, FACTS devices, HVDC systems, and Type-3 wind turbine generators will be investigated. These technologies---especially when realized using modular multilevel converters (MMCs)---introduce additional internal dynamics, control layers, and cross-coupling mechanisms that may significantly reshape the geometry of the resulting stability manifolds. Their systematic integration will enable the methodology to capture heterogeneous converter interactions in future hybrid transmission systems. Another important research direction concerns the incorporation of electromagnetic transient (EMT) effects into the linearization-based framework. While the current approach relies on averaged models, high-frequency converter dynamics and modulation-induced phenomena may influence system-level stability in converter-dominated grids. Future developments will therefore explore multi-frequency averaging and enhanced modeling techniques capable of embedding selected EMT-level dynamics into computationally tractable linear representations. Finally, the scalability of the proposed methodology to larger transmission networks will be investigated. Application to systems with increased size, geographical spread, and converter penetration will require computational enhancements, including structure-exploiting linearization, efficient eigenvalue screening, and advanced sampling strategies. These developments will support the use of stability manifolds in large-scale planning and interconnection studies.

\bibliographystyle{IEEEtran}

\end{document}